\documentclass[a4paper,12pt]{article}
\usepackage{amsmath}
\usepackage{amssymb}
\usepackage{latexsym}
\usepackage{enumerate}
\newcommand{\be}{\begin{eqnarray}}
\newcommand{\ee}{\end{eqnarray}}

\newcommand{\bsub}{\begin{subequations}}
\newcommand{\esub}{\end{subequations}}

\newcommand{\disfrac}[1][2]{\displaystyle\frac}
\begin{document}
\title{{\bf Towards Canonical Quantum Gravity for $G_1$ Geometries in 2+1
Dimensions with a $\Lambda$--Term}} \vspace{1cm}
\author{\textbf{T. Christodoulakis}\thanks{tchris@phys.uoa.gr}\,, \textbf{G. Doulis}\thanks{gdoulis@phys.uoa.gr}\,,
\textbf{Petros A. Terzis}\thanks{pterzis@phys.uoa.gr}\\
{\it Nuclear and Particle Physics Section, Physics Department,}\\{\it University of Athens, GR 157--71 Athens}\\
\textbf{E. Melas}\thanks{evangelosmelas@yahoo.co.uk}\\
{\it Technological Educational Institution of Lamia}\\
{\it Electrical Engineering Department, GR 35--100, Lamia }\\
\textbf{Th. Grammenos}\thanks{thgramme@uth.gr}\\
{\it Department of Mechanical Engineering, University of Thessaly,}\\
{\it GR 383--34 Volos}\\
\textbf{G.O. Papadopoulos}\thanks{gopapad@mathstat.dal.ca}\\
{\it Department of Mathematics and Statistics, Dalhousie University}\\
{\it Halifax, Nova Scotia, Canada B3H 3J5}\\
\textbf{A. Spanou}\thanks{aspanou@central.ntua.gr}\\
{\it School of Applied Mathematics and Physical Sciences},\\
{\it National Technical University of Athens, GR 157--80, Athens}}
\date{}
\maketitle
\begin{center}
\textit{}
\end{center}
\vspace{-1cm}
\newpage
\begin{abstract}
The canonical analysis and subsequent quantization of the
(2+1)-dimensional action of pure gravity plus a cosmological
constant term is considered, under the assumption of the existence
of one spacelike Killing vector field. The proper imposition of the
quantum analogues of the two linear (momentum) constraints reduces
an initial collection of state vectors, consisting of all smooth
functionals of the components (and/or their derivatives) of the
spatial metric, to particular scalar smooth functionals. The demand
that the midi-superspace metric (inferred from the kinetic part of
the quadratic (Hamiltonian) constraint) must define on the space of
these states an induced metric whose components are given in terms
of the same states, which is made possible through an appropriate
re-normalization assumption, severely reduces the possible state
vectors to three unique (up to general coordinate transformations)
smooth scalar functionals. The quantum analogue of the Hamiltonian
constraint produces a Wheeler-DeWitt equation based on this reduced
manifold of states, which is completely integrated.
\\
\\
{\bf PACS Numbers}: 04.60.Ds, 04.60.Kz
\end{abstract}

\numberwithin{equation}{section}
\section{Introduction}

Dirac's seminal work on his formalism for a self-contained treatment
of systems with constraints \cite{Dirac1}, \cite{Dirac2},
\cite{Dirac3}, \cite{Dirac} has paved the way for a systematic
treatment of constrained dynamics. Some of the landmarks in the
study of constrained systems have been the connection between
constraints and invariances \cite{Bergman}, the extension of the
formalism to describe fields with half-integer spin through the
algebra of Grassmann variables \cite{Berezin} and the introduction
of the BRST formalism \cite{Becchi}. All the classical results
obtained so far have made up an armory prerequisite for the
quantization of gauge theories and there are several excellent
reviews studying constraint systems with a finite number of degrees
of freedom \cite{Sudarshan} or constraint field theories
\cite{Hanson}, as well as more general presentations
\cite{Sundermeyer}, \cite{Gitman}, \cite{Govaers}, \cite{Henneaux},
\cite{Wipf}, \cite{Thiemann}. In particular, the conventional
canonical analysis approach of quantum gravity has been initiated by
B.S. DeWitt \cite{DeWitt} based on earlier work of P.G. Bergmann
\cite{Bergmann}.

In the absence of a full theory of quantum gravity, it is reasonably
important to address the quantization of (classes of) simplified
geometries. An elegant way to achieve a degree of simplification is
to impose some symmetry. For example, the assumption of a $G_3$
symmetry group acting simply transitively on the surfaces of
simultaneity, i.e. the existence of three independent space-like
Killing vector fields, leads to classical and subsequently quantum
homogeneous cosmology (see, e.g., \cite{Ryan}, \cite{chris1}). The
imposition of lesser symmetry, e.g. fewer Killing vector fields,
results in the various inhomogeneous cosmological models
\cite{Krasin}. The canonical analysis under the assumption of
spherical symmetry, which is a $G_3$ group acting multiply
transitively on two-dimensional space-like subsurfaces of the
three-slices, has been first considered in \cite{Thomi},
\cite{Hajic}. Quantum black holes have also been treated, for
instance, in \cite{Zan}, \cite{Kief1} while in \cite{Kief} a lattice
regularization has been employed to deal with the infinities arising
due to the ill-defined nature of the quantum operator constraints.

Another way to arrive at simplified models is to consider lower
dimensions. For example, there is a vast literature on
(2+1)-dimensional gravity (see, e.g., \cite{Carlip1},
\cite{Carlip2}, \cite{Carlip3}  and references there in). The role
of non-commutative geometry in (2+1)-dimensional quantum gravity has
been recently investigated in \cite{schro}. In this work we consider
the canonical quantization of all 2+1 geometries admitting one
spacelike Killing vector field. In Section 2 we give the reduced
metrics, the space of classical solutions and the Hamiltonian
formulation of the reduced Einstein-Hilbert action principle,
resulting in one (quadratic) Hamiltonian and two (linear) momentum
first class constraints. In Section 3 we consider the quantization
of this constrained system following Dirac's proposal of implementing
the quantum operator constraints as conditions annihilating the
wave-function \cite{Dirac}. Our guide-line is a conceptual
generalization of the quantization scheme developed in
\cite{Kuchar1}, \cite{Kuchar2} for the case of constrained systems
with finite degrees of freedom, to the present case. Even though
after the symmetry reduction the system still represents a field
theory (all remaining metric components depend on time and the
radial coordinate), we manage to extract and subsequently completely
solve a Wheeler-DeWitt equation in terms of three unique smooth
scalar functionals of the appropriate components of the reduced
spatial metric. This is achieved through an appropriate
re-normalization assumption we adopt. Finally, some concluding
remarks are included in the discussion.

\section{Possible Metrics and Hamiltonian Formulation}

Our starting point is the action principle:
\begin{equation}\label{Action}
I=\int d^3x\sqrt{-g}(R-2\Lambda).
\end{equation}
The equations of motion arising upon variation of this action are
\begin{equation}\label{eqmotion}
R_{IJ}-\frac{1}{2}g_{IJ}R+\Lambda g_{IJ}=0,
\end{equation}
where $I,J=0,1,2$. Of course, since in three dimensions the Riemann
curvature tensor is expressible in terms of both the Ricci tensor
and scalar, the space of solutions to (\ref{eqmotion}) consists
simply of all maximally symmetric 3D metrics (AdS3). If topological
considerations are taken into account, the above space might be
``enriched" containing, for example, the stationary BTZ ``black''
hole \cite{Zanelli1}, \cite{Zanelli2}
\begin{equation}\label{BTZ}
ds^2=-(M-\Lambda r^2)dt^2-Jdtd\phi+\left(M-\Lambda
r^2+\frac{J^2}{4r^2}\right)^{-1}dr^2+r^2d\phi^2
\end{equation}
or the ``cosmological'' solutions \cite{Pap}, \cite{Tsagas}
\begin{eqnarray}
\label{cosmo1}ds^2&=&-\frac{1}{4t^2\Lambda}dt^2+\frac{1}{2t\sqrt{\Lambda}}(dr^2+d\phi^2) ,\\
\label{cosmo2}ds^2&=&-\left(\frac{4}{16t^2-\Lambda}\right)^2dt^2+\frac{4}{16t^2-\Lambda}dr^2+\frac{4e^{-4r}}{16t^2-\Lambda}d\phi^2.
\end{eqnarray}
Note that all these three line elements are locally AdS3 and
therefore admit six local Killing fields. Their differences consist
in the topological identifications. At this point, we deem it
pertinent to explain our view concerning the issue of the bearing of
topology on a local theory: The Hamiltonian formulation is by itself
implying a space-time topology $R \times \Sigma^2$. Consequently,
what we are concerned with is the topology of the 2-slices. Since
the theory is local, it is implicitly assumed that the entire
analysis holds in a coordinate patch. Different topologies can only
affect the number of patches needed to cover the space and,
therefore, can only impose restrictions on the range of the
coordinates and/or the range of validity of local fields, such as
the symmetry generators admitted by these metrics; The paradigm of
the cylinder may help clarify our point: The integral curves of
rotations in the plane are circles, but if one tries to draw a
circle of radius $R \geq 2 \pi L$ on the cylinder (L being the
cylinder's radius), crossings (or a pinch in case of equality) will
occur, indicating that the corresponding generator is ill-defined.
In such a situation one can, as many do, drop rotations altogether;
this is the case in \cite{Zanelli1}, \cite{Zanelli2}, where four of
the six Killing fields are considered as non-valid symmetries. On
the other hand one can accept integral curves (circles) of radius
$R<2 \pi L$ (by suitably restricting the range of validity of the
Killing field), which would simply result in the need of two patches
to cover the cylinder with these lines. We adopt this latter point
of view, as it seems to us much more reasonable. We shall thus not
specify any ranges for our coordinates $(t,r,\phi)$ precisely to
allow for different topological options, which are not otherwise
affecting our results.\\
In this spirit we can say that the above metrics admit a $G_6$
symmetry group. In what follows, we consider a generalization
consisting in the imposition of a $G_1$ symmetry only, i.e we impose
one Killing vector field, say $\xi=\frac{\partial}{\partial \phi}$.
Subsequently, all components of the metric become functions of both
the time and the radial coordinate only. The canonical decomposition
of such a metric is given in terms of the spatial metric
$g_{ij}(t,r)$, the lapse function $N^o(t,r)$ and the shift
``vector'' $N_i(t,r)$ \cite{Sundermeyer}:
\begin{equation}\label{canmetric}
ds^2=\left(-(N^o)^2+g^{ij}N_iN_j\right)dt^2+2N_idtdx^i+g_{ij}dx^idx^j,
\end{equation}
where
\begin{equation}\label{g metric}
g_{ij}=
\begin{pmatrix}
\displaystyle{\frac{\rho^2+\sigma^2\chi^2}{\sigma}} &
\displaystyle{\sigma\chi} \cr & \cr \displaystyle{\sigma\chi} &
\displaystyle{\sigma}
\end{pmatrix}, \qquad
g^{ij}=
\begin{pmatrix}
\displaystyle{\frac{\sigma}{\rho^2}} &
-\displaystyle{\frac{\sigma\chi}{\rho^2}} \cr & \cr
-\displaystyle{\frac{\sigma\chi}{\rho^2}} &
\displaystyle{\frac{\rho^2+\sigma^2\chi^2}{\sigma\rho^2}}
\end{pmatrix}
\end{equation}
with $i,j=1,2$, and $x^i=(r,\phi)$. The particular parametrization
of $g_{ij}$ above has been chosen in such a way as to simplify the
second linear constraint (see below), and consequently the resulting
algebra.

For the Hamiltonian formulation of the system (\ref{canmetric})
(see, e.g., chapter 9 of \cite{Sundermeyer}), we first define the
vectors
\[\eta^I=\frac{1}{N^o}\left(1,-N^i\right), \qquad N^i\equiv g^{ik}N_k\]
\[\,F^I=\eta^J_{;J}\,\eta^I-\eta^I_{;J}\,\eta^J\]
where $I$, $J$ are space-time indices and ``\,$;$\," stands for
covariant differentiation with respect to (\ref{canmetric}). Then,
utilizing the Gauss-Codazzi equation (see, e.g., \cite{eisen}), we
eliminate all second time-derivatives  from the Einstein-Hilbert
action and arrive at an action quadratic in the velocities, $I=\int
d^3x\sqrt{-g}(R-2\Lambda-2\,F^I_{;I})$. The application of the Dirac
algorithm results firstly in the three primary constraints
$P_o\equiv\frac{\delta
L}{\delta\dot{N}^o}\approx0,\,P^i\equiv\frac{\delta
L}{\delta\dot{N}_i}\approx0$ and the Hamiltonian
\begin{equation}\label{Hamiltonian}
H=\int \left(N^o\mathcal{H}_o+N^i\,\mathcal{H}_i\,\right)dr,
\end{equation}
where $\mathcal{H}_o$, $\mathcal{H}_i$ are given by
\bsub\label{constraints}
\begin{eqnarray}
\label{Ho}\mathcal{H}_o &=& \frac{1}{2}\,G^{\alpha\beta}\,\pi_\alpha\,\pi_\beta+V\\
\label{H1}\mathcal{H}_1 &=& \sigma^\prime\,\pi_\sigma-\rho\,\pi_\rho^\prime-\chi\,\pi_\chi^\prime\\
\label{H2}\mathcal{H}_2 &=&-\pi_\chi^\prime,
\end{eqnarray}
\esub the indices $(\alpha,\beta)$ take the values
$(\rho,\sigma,\chi)$ and
$^{\prime}\equiv\displaystyle{\frac{\partial}{\partial r}}$. The
Wheeler-DeWitt midi-superspace metric $G^{\alpha\beta}$ reads
\begin{equation}\label{supermetric}
G^{\alpha\beta}=
\begin{pmatrix}
-\displaystyle{\rho} & -\displaystyle{\sigma} & 0 \cr & \cr
-\displaystyle{\sigma} & 0 & 0 \cr & \cr 0 & 0 &
\displaystyle{\frac{\rho}{\sigma^2}}
\end{pmatrix},
\end{equation}
while the potential $V$ is
\begin{equation}\label{potential}
V=2\Lambda\rho+\left(\frac{\sigma^\prime}{\rho}\right)^\prime.
\end{equation}
The requirement for preservation, in time, of the primary
constraints leads to the secondary constraints
\begin{equation}\label{conweak}
\mathcal{H}_o\approx0,\qquad\mathcal{H}_1\approx0,\qquad
\mathcal{H}_2\approx0.
\end{equation}
At this stage, a tedious but straightforward calculation produces
the following ``open'' Poisson bracket algebra of these constraints:
\be\label{alge}
\{\mathcal{H}_o(r),\mathcal{H}_o(\tilde{r})\}&=&[g^{1j}(r)\mathcal{H}_j(r)+g^{1j}(\tilde{r})\mathcal{H}_j(\tilde{r})]\delta'(r,\tilde{r})\nonumber\\
\{\mathcal{H}_1(r),\mathcal{H}_o(\tilde{r})\}&=&\mathcal{H}_o(r)\delta'(r,\tilde{r})\nonumber\\
\{\mathcal{H}_2(r),\mathcal{H}_o(\tilde{r})\}&=&0\\
\{\mathcal{H}_1(r),\mathcal{H}_1(\tilde{r})\}&=&\mathcal{H}_1(r)\delta'(r,\tilde{r})-\mathcal{H}_1(\tilde{r})\delta(r,\tilde{r})'\nonumber\\
\{\mathcal{H}_1(r),\mathcal{H}_2(\tilde{r})\}&=&\mathcal{H}_2(r)\delta'(r,\tilde{r})\nonumber\\
\{\mathcal{H}_2(r),\mathcal{H}_2(\tilde{r})\}&=&0\nonumber \ee
indicating that they are first class and also signaling the
termination of the algorithm. Thus, our system is described by
(\ref{conweak}); the ``dynamical'' Hamilton-Jacobi equations
$\disfrac{d\,\pi_\rho}{d\,t}=\{\pi_\rho,H\}$,
$\disfrac{d\,\pi_\sigma}{d\,t}=\{\pi_\sigma,H\}$,
$\disfrac{d\,\pi_\chi}{d\,t}=\{\pi_\chi,H\}$ are satisfied by virtue
of the time derivatives of (\ref{conweak}). One can readily check
(as one must always do with reduced action principles) that these
three equations, when expressed in the velocity phase-space with the
help of the definitions $\disfrac{d\,\rho}{d\,t}=\{\rho,H\}$,
$\disfrac{d\,\sigma}{d\,t}=\{\sigma,H\}$,
$\disfrac{d\,\chi}{d\,t}=\{\chi,H\}$, are completely equivalent to
the three independent Einstein's field equations satisfied by
(\ref{canmetric}).

We end up this section by noting a few facts concerning the
transformation properties of $\rho(t,r),\, \sigma(t,r),\, \chi(t,r)$
and their spatial derivatives under changes of the radial variable
$r$ of the form $r\rightarrow\tilde{r}=h(r)$. As it can easily be
inferred from (\ref{canmetric}) and (\ref{g metric}):
\begin{equation}\label{tensor}
  \begin{split}
    & \tilde{\rho}(\tilde{r})=\rho(r)\,\frac{d\,r}{d\,\tilde{r}}, \qquad \tilde{\sigma}(\tilde{r})=\sigma(r),  \qquad \tilde{\chi}(\tilde{r})=\chi(r)\,\frac{d\,r}{d\,\tilde{r}},\\
    & \frac{d\,\tilde{\sigma}(\tilde{r})}{d\,\tilde{r}}=\frac{d\,\sigma(r)}{d\,r}\,\frac{d\,r}{d\,\tilde{r}}, \qquad \frac{d}{d\,\tilde{r}}\left(\frac{\tilde{\chi}(\tilde{r})}{\tilde{\rho}(\tilde{r})}\right)=\frac{d}{d\,r}\left(\frac{\chi(r)}{\rho(r)}\right)\,\frac{d\,r}{d\,\tilde{r}},
  \end{split}
\end{equation}
where the $t$-dependence has been omitted for the sake of brevity.
Thus, under the above coordinate transformations,
$\sigma,\,\frac{\chi}{\rho}$ are scalars, while $\rho,\,\chi$ and
the derivatives of $\sigma,\,\frac{\chi}{\rho}$ are covariant rank 1
tensors (one-forms), or, equivalently in one dimension, scalar
densities of weight $-1$. Therefore, the scalar derivative is not
$\disfrac{d}{d\,r}$ but rather $\disfrac{d}{\rho\,d\,r}$ or
$\disfrac{d}{\chi\,d\,r}\equiv\frac{\rho}{\chi}\disfrac{d}{\rho\,d\,r}$.
Finally, if we consider an infinitesimal transformation
$r\rightarrow\tilde{r}=r-\eta(r)$, it is easily seen that the
corresponding changes induced on the basic fields are:
\begin{equation}\label{lieder}
\delta\,\rho(r)=(\rho(r)\,\eta(r))^{\prime}, \qquad
\delta\,\sigma(r)=\sigma^{\prime}(r)\,\eta(r), \qquad
\delta\,\chi(r)=(\chi(r)\,\eta(r))^{\prime}
\end{equation}
i.e., nothing but the one-dimensional analogue of the appropriate
Lie derivatives.

With the use of (\ref{lieder}), we can reveal the nature of the
action of $\mathcal{H}_1$ on the basic configuration space variables
as that of the generator of spatial diffeomorphisms:
\begin{equation}\label{sdif}
\begin{split}
& \left\{\rho(r)\,,\,\int d\tilde{r}\,\eta(\tilde{r})\,\mathcal{H}_1(\tilde{r})\right\}=(\rho(r)\,\eta(r))^{\prime},\\
& \left\{\sigma(r)\,,\,\int
d\tilde{r}\,\eta(\tilde{r})\,\mathcal{H}_1(\tilde{r})\right\}=\sigma^{\prime}(r)\,\eta(r),\\
& \left\{\chi(r)\,,\,\int
d\tilde{r}\,\eta(\tilde{r})\,\mathcal{H}_1(\tilde{r})\right\}=(\chi(r)\,\eta(r))^{\prime}.
\end{split}
\end{equation}
Thus, we are justified to consider $\mathcal{H}_1$ as the
representative, in phase-space, of an arbitrary infinitesimal
reparametrization of the radial coordinate. As far as
$\mathcal{H}_2$ is concerned, the situation is a little more
complicated: the imposition of the symmetry generated by the Killing
vector field $\xi=\partial/\partial \phi$ has left all configuration
variables without any $\phi$ dependence; subsequently we can not
expect $\mathcal{H}_2$ to generate arbitrary infinitesimal
reparametrization of $\phi$. Nevertheless, we can identify a
property of $\mathcal{H}_2$ which links its existence to the
existence of $\xi$. This property is described by the relation:
$\{\mathcal{H}_2(r),\{\mathcal{H}_2(r),\{\mathcal{H}_2(r),g_{ij}(\tilde{r})\}\}\}=0
\underset{ndence}{\overset{correspo}{\Longleftrightarrow}}
\mathcal{L}_\xi g_{ij}=0$.

\section{Quantization}
We are now interested in attempting to quantize this Hamiltonian
system following Dirac's general spirit of realizing all the classical
first class constraints (\ref{conweak}) as quantum operator
constraint conditions annihilating the wave functional. The main
motivation behind such an approach is the justified desire to
construct a quantum theory manifestly invariant under the ``gauge"
generated by the constraints. To begin with, let us first note that,
despite the simplification brought by the imposition of the symmetry
$\xi=\partial/\partial \phi\Leftrightarrow \mathcal{L}_\xi
g_{IJ}=0$, the system  is still a field theory in the sense that all
configuration variables and canonical conjugate momenta depend not
only on time (as is the case in homogeneous cosmology), but also on
the radial coordinate $r$. Thus, to canonically quantize the system
in the Schr\"{o}dinger representation, we first realize the
classical momenta as functional derivatives with respect to their
corresponding conjugate fields
\[\hat{\pi}_{\rho}(r)=-i\,\disfrac{\delta}{\delta\,\rho(r)}, \qquad\hat{\pi}_{\sigma}(r)=-i\,
\disfrac{\delta}{\delta\,\sigma(r)},
\qquad\hat{\pi}_{\chi}(r)=-i\,\disfrac{\delta}{\delta\,\chi(r)}.\]

We next have to decide on the initial space of state vectors. To
elucidate our choice, let us consider the action of a momentum
operator on some function of the configuration field variables, say
\[\hat{\pi}_{\rho}(r) \rho(\tilde{r})^2=-2i\rho(\tilde{r}) \delta(\tilde{r},r).\]
The Dirac delta-function renders the outcome of this action a
distribution rather than a function. Also, if the momentum operator
were to act at the point at which the function is evaluated, i.e. if
$\tilde{r}=r$, then its action would produce a $\delta(0)$ and would
therefore be ill-defined. Both of these unwanted features are
rectified, as far as expressions linear in momentum operators are
concerned, if we choose as our initial collection of states all
$\emph{smooth functionals}$ (i.e., integrals over $r$) of the
configuration variables $\rho(r),\sigma(r),\chi(r)$ and their
derivatives of any order. Indeed, as we infer from the previous
example,
\[\hat{\pi}_{\rho}(r)\int d\tilde{r} \rho(\tilde{r})^2=
-2i\,\int d\tilde{r}\rho(\tilde{r})
\delta(\tilde{r},r)=-2i\rho(r);\] thus the action of the momentum
operators on all such states will be well-defined (no $\delta(0)$'s)
and will also produce only local functions and not distributions.
However, even so, $\delta(0)$'s will appear as soon as local
expressions quadratic in  momenta are considered, e.g.,
\[\hat{\pi}_{\rho}(r)\,\hat{\pi}_{\rho}(r)\int d\tilde{r}
\rho(\tilde{r})^2=\hat{\pi}_{\rho}(r) (-2i\int
d\tilde{r}\rho(\tilde{r})
\delta(\tilde{r},r))=\hat{\pi}_{\rho}(r)(-2i\rho(r))=-2i
\delta(r,r).\]

Another problem of equal, if not greater, importance has to do with
the number of derivatives (with respect to $r$) considered: A
momentum operator acting on a smooth functional of degree $\emph{n}$
in derivatives of $\rho(r),\sigma(r),\chi(r)$ will, in general,
produce a function of degree $2n$, e.g.,
\[\hat{\pi}_{\rho}(r)\int d\tilde{r}
\rho''(\tilde{r})^2=-2i\int d\tilde{r}
\rho''(\tilde{r})\delta''(\tilde{r},r)=-2i\rho^{(4)}(r).\] Thus,
clearly, more and more derivatives must be included if we desire the
action of momentum operators to keep us inside the space of
integrands corresponding to the initial collection of smooth
functionals; eventually, we have to consider $n\rightarrow\infty$.
This, in a sense, can be considered as the reflection to the
canonical approach, of the non-re-normalizability results existing
in the so-called covariant approach. The way to deal with these
problems is, loosely speaking, to regularize (i.e., render finite)
the infinite distribution limits, and re-normalize the theory by,
 somehow, enforcing n to terminate at some finite value.

In the following, we are going to present a quantization scheme of
our system which: (a) avoids the occurrence of $\delta(0)$'s, (b)
reveals the value $n=1$, as the only possibility to obtain a closed
space of state vectors, and (c) extracts a finite-dimensional
Wheeler-DeWitt equation governing the quantum dynamics. The scheme
closely parallels, conceptually, the quantization developed in
\cite{Kuchar1},\cite{Kuchar2} for finite systems with one quadratic
and a number of linear first class constraints. Therefore, we deem
it appropriate and instructive to present a brief account of the
essentials of this construction.

To this end, let us consider a system described by a Hamiltonian of
the form \be
H & \equiv & \mu X+\mu^i \chi_i \nonumber\\
& = &\mu\,\left(\frac{1}{2}G^{AB}(Q^\Gamma)P_AP_B\,+U^A(Q^\Gamma)
P_A+V(Q^\Gamma)\right)\, +\mu^i\,\phi_i^A(Q^\Gamma)P_A, \ee where
$A,B,\Gamma\ldots=1,2\ldots, M$ count the configuration space
variables and\break $i=1,2,\ldots, N<(M-1)$ numbers the
super-momenta constraints $\chi_i\approx0$, which along with the
super-Hamiltonian constraint $X\approx0$ are assumed to be first
class:
\begin{equation}\label{algekuch}
\{X,X\}=0, \qquad \{X,\chi_i\}=X C_i+C^j_i \chi_j, \qquad
\{\chi_i,\chi_j\}=C^k_{ij} \chi_k,
\end{equation}
where the first (trivial) Poisson bracket has been included only to
emphasize the difference from the first of (\ref{alge}).

The physical state of the system is unaffected by the ``gauge"
transformations generated by $(X,\,\chi_i)$, {\bf but also} under
the following three changes:
\begin{enumerate}[(I)]
\item Mixing of the super-momenta with a non-singular matrix
\[\bar{\chi}_i=\lambda^j_i(Q^\Gamma)  \chi_j\]
\item Gauging of the  super-Hamiltonian with the super-momenta
\[\bar{X}=X+\kappa^{(Ai}(Q^\Gamma)\phi_i^{B)}(Q^\Gamma)P_AP_B+\sigma^i(Q^\Gamma)\phi_i^A(Q^\Gamma)P_A\]
\item Scaling of the super-Hamiltonian
\[\bar{X}=\tau^2(Q^\Gamma)X\]
\end{enumerate}
Therefore, the geometrical structures on the configuration space
that can be inferred from the super-Hamiltonian are really
equivalence classes under actions (I), (II) and (III); for example
(II), (III) imply that the super-metric $G^{AB}$ is known only up to
conformal scalings and additions of the super-momenta coefficients
$\bar{G}^{AB}=\tau^2(G^{AB}+\kappa^{(Ai}\phi_i^{B)})$. It is thus
mandatory that, when we Dirac-quantize the system, we realize the
quantum operator constraint conditions on the wave-function in such
a way as to secure that the whole scheme is independent of actions
(I), (II), (III). This is achieved by the following steps:
\begin{enumerate}[(1)]
\item Realize the linear operator constraint conditions with the momentum
operators to the right
\[\hat{\chi}_i\Psi=0\leftrightarrow\phi_i^A(Q^\Gamma)\frac{\partial\,\Psi(Q^\Gamma)}{\partial\,Q^A}=0,\]
which maintains the geometrical meaning of the linear constraints
and produces the $M-N$ independent solutions to the above equations
$q^\alpha(Q^\Gamma),\,\alpha=1,2,\ldots, M-N$ called physical
variables, since they are invariant under the transformations
generated by $\hat{\chi}_i$.
\item  In order to make the final states physical with respect to the
``gauge" generated by the quadratic constraint $\hat{X}$ as well:

Define the induced structure $g^{\alpha\beta}\equiv
G^{AB}\disfrac{\partial\,q^\alpha}{\partial\,Q^A}\disfrac{\partial\,q^\beta}{\partial\,Q^B}$
and realize the quadratic in momenta part of $X$ as the conformal
Laplace-Beltrami operator based on $g_{\alpha\beta}$. Note that in
order for this construction to be self consistent, all components of
$g_{\alpha\beta}$ must be functions of the physical coordinates
$q^\gamma$. This can be proven to be so by virtue of the classical
algebra the constraints satisfy (for specific quantum cosmology
examples see \cite{chris1}).
\end{enumerate}

We are now ready to proceed with the quantization of our system, in
close analogy to the scheme above outlined. In order to realize the
equivalent to step 1, we first define the quantum analogue of
$\mathcal{H}_1(r)\approx 0$ as
\begin{equation}\label{qulincon}
\hat{\mathcal{H}}_1(r)\Phi=0\leftrightarrow
-\rho(r)\,(\frac{\delta\,\Phi}{\delta\,\rho(r)})'+\sigma'(r)\,\frac{\delta\,\Phi}{\delta\,\sigma(r)}-
\chi(r)\,(\frac{\delta\,\Phi}{\delta\,\chi(r)})'=0.
\end{equation}
As explained in the beginning of the section, the action of
$\hat{\mathcal{H}}_1(r)$ on all smooth functionals is well defined,
i.e., produces no $\delta(0)$'s. It can be proven that, in order for
such a functional to be annihilated by this linear quantum operator,
it must be scalar, i.e. have the form \bsub\label{scalfun}
\begin{eqnarray}
\label{Phi1}\Phi&=&\int\rho(\tilde{r})\,f\left(\Sigma^{(0)},\Sigma^{(1)},\ldots,\Sigma^{(n)},X^{(0)},X^{(1)},
\ldots,X^{(n)}\right)d\tilde{r}\\
\label{Sigma}\Sigma^{(0)}&\equiv&\sigma(\tilde{r}),\quad\Sigma^{(1)}\equiv\frac{\sigma'(\tilde{r})}{\rho(\tilde{r})},
\ldots,\quad\Sigma^{(n)}\equiv\frac{1}{\rho(\tilde{r})}\frac{d}{d\tilde{r}}
\left(\underset{n-1}{\underbrace{\ldots}}\,\,\sigma(\tilde{r})\right)\\
\label{X}X^{(0)}&\equiv&\frac{\chi(\tilde{r})}{\rho(\tilde{r})},\quad
X^{(1)}\equiv\frac{1}{\rho(\tilde{r})}
\left(\frac{\chi(\tilde{r})}{\rho(\tilde{r})}\right)',\ldots,\quad
X^{(n)}\equiv\frac{1}{\rho(\tilde{r})}
\frac{d}{d\tilde{r}}\left(\underset{n-1}{\underbrace{\ldots}}\,\,\frac{\chi(\tilde{r})}{\rho(\tilde{r})}\right)
\end{eqnarray}
\esub where $f$ is any function of its arguments. We note that, as
it is discussed at the end of the previous section,
$\frac{\sigma'}{\rho}$ is the only scalar first derivative of
$\sigma$, and likewise for the higher derivatives. The proof of this
statement is analogous to the proof of the corresponding result
concerning full gravity \cite{Tracy}: consider an infinitesimal
$r$-reparametrization $\tilde{r}=r-\eta(r)$. Under such a change,
the left-hand side of (\ref{scalfun}), being a number, must remain
unaltered. If we calculate the change induced on the right-hand side
we arrive at
\begin{equation}
0=\int\left[f \delta \rho+\rho \frac{\delta f}{\delta \sigma}\delta
\sigma +\rho\frac{\delta f}{\delta (\chi/\rho)}\delta
\left(\frac{\chi}{\rho}\right)\right]dr= \int [\rho
\hat{\mathcal{H}}_1 (f)]\eta(r)dr,
\end{equation}
where use of (\ref{lieder}) and a partial integration has been made.
Since this must hold for any $\eta(r)$, the result sought for is
obtained.

We now turn to the second linear constraint and try to see what are
the restrictions it brings into our space of state vectors. We
define
\begin{equation}\label{qulincon2}
\hat{\mathcal{H}}_2(r)\Phi=0\leftrightarrow(\frac{\delta\,\Phi}
{\delta\,\chi(r)})'=0\leftrightarrow\frac{\delta\,\Phi}{\delta\,\chi(r)}=k,
\end{equation}
where k is any constant (with respect to r) independent of the basic
fields and their derivatives, and $\Phi$ is given by
(\ref{Phi1})$-$(\ref{X}). As we argued before, the functional
derivative $\frac{\delta}{\delta\chi(r)}$ acting on $X^{(n)}$ will
produce, upon partial integration of the $n^{th}$ derivative of the
Dirac delta function, a term proportional to $X^{(2n)}$. Since the
arguments of $f$ in (\ref{Phi1}) reach only up to $X^{(n)}$, it is
evident that $f$ must be such that the coefficient of $X^{(2n)}$
vanishes; more precisely
\begin{eqnarray*}
\frac{\delta\,\Phi}{\delta\,\chi(r)}=k\leftrightarrow &\ldots&+\int
\rho(\tilde{r})\frac{\partial f}{\partial X^{(n)}(\tilde{r})}
\frac{\delta X^{(n)}(\tilde{r})}{\delta \chi(r)}\,d\tilde{r}=k\leftrightarrow\\
&\ldots&+\int \rho(\tilde{r})\frac{\partial f}{\partial
X^{(n)}(\tilde{r})}
\frac{1}{\rho(\tilde{r})}\frac{d}{d\tilde{r}}\left(\underset{n-1}{\underbrace{\ldots}}\,\,
\frac{\delta(r,\tilde{r})}{\rho(\tilde{r})}\right)d\tilde{r}=k\leftrightarrow\\
&\ldots&+(-1)^n\int \frac{\partial^2 f}{\partial
(X^{(n)}(\tilde{r}))^2}\,
X^{(2n)}(\tilde{r})\delta(r,\tilde{r})\,d\tilde{r}=k\leftrightarrow\\
&\ldots&+(-1)^n\frac{\partial^2 f}{\partial
(X^{(n)})^2}\,X^{(2n)}=k.
\end{eqnarray*}
Thus, since all the terms hidden in $\ldots$ do not involve
$X^{(2n)}$ and \eqref{qulincon2} must be satisfied identically for
all $X^k$'s $\,k=0,1,...2n$ , we conclude that $\frac{\partial^2
f}{\partial (X^{(n)})^2}=0$ in order for this equation to have a
possibility to be satisfied. Subsequently:
\begin{eqnarray*}
f=f_1\left(\Sigma^{(0)},\ldots,\Sigma^{(n)},X^{(0)},\ldots,X^{(n-1)}\right)X^{(n)}+
f_2\left(\Sigma^{(0)},\ldots,\Sigma^{(n)},X^{(0)},\ldots,X^{(n-1)}\right).
\end{eqnarray*}
Now, the term in $\Phi$ corresponding to $f_1$ is, up to a surface
term, equivalent to a general term depending on
$X^{(0)},\ldots,X^{(n-1)}$ only: indeed,
\begin{eqnarray*}
\Phi_1=\int
\rho(\tilde{r})f_1\frac{1}{\rho(\tilde{r})}\frac{d}{d\tilde{r}}X^{(n-1)}d\tilde{r},
\end{eqnarray*}
which upon subtraction of the surface term
\begin{eqnarray*}
A=\int d\tilde{r}\frac{d}{d\tilde{r}}\left(\int dX^{(n-1)}f_1\right)
\end{eqnarray*}
produces a smooth functional with arguments up to $X^{(n-1)}$ only.
Since a surface term in $\Phi$ does not affect the outcome of the
variational derivative $\frac{\delta\,\Phi}{\delta\,\chi(r)}$, we
conclude that only $f_2$ is important for the local part of $\Phi$.
The entire argument can be repeated successively for
$n-1,\,n-2,\ldots,\,1$; Therefore all $X^{(n)}$'s are suppressed
from $f$ except when $n=0$. Thus, finally, upon inserting into
(\ref{qulincon2}) the resulting functional:
\begin{eqnarray*}
\Phi&=&\int\rho(\tilde{r})\,h\left(\Sigma^{(0)},\ldots,\Sigma^{(n)},X^{(0)}
\right)d\tilde{r}
\end{eqnarray*}
we obtain
\begin{eqnarray*}
\frac{\delta\,\Phi}{\delta\,\chi(r)}&=&k\leftrightarrow \int
\rho(\tilde{r})\frac{\partial h}{\partial X^{(0)}}
\frac{\delta(r,\tilde{r})}{\rho(\tilde{r})}d\tilde{r}=k\leftrightarrow
\frac{\partial h}{\partial X^{(0)}}=k\leftrightarrow\\
h&=&k\,\frac{\chi(r)}{\rho(r)}+L\left(\Sigma^{(0)},\ldots,\Sigma^{(n)}\right).
\end{eqnarray*}
We have thus reached the conclusion that the imposition of both
linear quantum operators $\hat{\mathcal{H}}_1$ and
$\hat{\mathcal{H}}_2$ dictates the form of the smooth functional to
be:
\begin{equation}\label{Phi2}
\Phi=k\int d\tilde{r}\,\chi(\tilde{r})+\int
d\tilde{r}\,\rho(\tilde{r})\,L
\left(\Sigma^{(0)},\ldots,\Sigma^{(n)}\right).
\end{equation}

We now try to realize step 2 of the programm previously outlined. We
have to define the equivalent of Kucha\v{r}'s induced metric on the
so far space of ``physical'' states $\Phi$ described by (\ref{Phi2})
which are the analogues, in our case, of Kucha\v{r}'s physical
variables $q^\alpha$. Let us start our investigation by considering
\textbf{one} initial candidate of the above form. Then, generalizing
the partial to functional derivatives, the induced metric will be
given by
\begin{equation}\label{physical}
g^{\Phi\Phi}=G^{\alpha\beta}\,\frac{\delta\,\Phi}{\delta\,x^\alpha}\,
\frac{\delta\,\Phi}{\delta\,x^\beta},
\end{equation}
where $(x^\alpha, x^\beta)=(\rho,\sigma,\chi)$ and $G^{\alpha\beta}$
is given by (\ref{supermetric}). Note that this metric is well
defined since it contains only first functional derivatives of the
state vectors, as opposed to any second order functional derivative
operator that might have been considered as a quantum analogue of
the kinetic part of $\mathcal{H}_o$. Nevertheless, $g^{\Phi\Phi}$
\emph{is} a local function and not a smooth functional. It is thus
clear that, if we want the induced metric $g^{\Phi\Phi}$ to be
composed out of the ``physical'' states annihilated by
$\hat{H}_1,\,\hat{H}_2$, we must establish a correspondence between
local functions and smooth functionals. A way
to achieve this is to adopt the following ansatz: \\
\\
\textbf{Assumption:} \emph{We assume that, as part of the
re-normalization procedure, we are permitted to map local functions
to their corresponding smeared expressions e.g.,
$\chi(r)\leftrightarrow \int d\tilde{r}\chi(\tilde{r})$.}\\
\\
Let us be more specific, concerning the meaning of the above
Assumption. Let $\mathcal{F}$ be the space which contains all local
functions, and define the equivalence relations
\begin{equation}\thicksim: \{f_1(r)\thicksim
f_2(\tilde{r}),\tilde{r}=g(r) \}, \quad \thickapprox:
\{h_1(r)\thickapprox
h_2(\tilde{r})\,\frac{d\,\tilde{r}}{d\,r},\tilde{r}=g(r)
\}\end{equation} for scalars and densities respectively.

Now let $\mathcal{F}_o=\{f\in \mathcal{F}, \mod
(\thicksim,\thickapprox)\}$ and $\mathcal{F_I}$ the space of the
smeared functionals. We define the one to one maps $\mathfrak{G}$,
$\mathfrak{G}^{-1}$
\begin{equation} \mathfrak{G}: \mathcal{F}_o \mapsto
\mathcal{F}_I: \quad \chi(r) \mapsto \int\chi(r)\,dr, \quad
\mathfrak{G}^{-1}: \mathcal{F}_I \mapsto \mathcal{F}_o: \quad
\int\chi(r)\,dr \mapsto \chi(r)
\end{equation}
\\
The necessity to define the maps $\mathfrak{G}, \mathfrak{G}^{-1}$
on the equivalence classes and not on the individual functions,
stems out of the fact that we are trying to develop a quantum theory
of the geometries \eqref{canmetric}, \eqref{g metric} and not of
their coordinate representations. If we had tried to define the map
$\mathfrak{G}$ from the original space $\mathcal{F}$ to
$\mathcal{F}_I$ we would end up with states which would not be
invariant under spatial coordinate transformations ($\mathbf{r}$ -
reparameterizations). Indeed, one can make a correspondence between
local functions and smeared expressions, but smeared expressions
\emph{must} contain another arbitrary smearing function, say $s(r)$.
Then the map between functions and smeared expressions is one to one
(as is also the above map) and is given by multiplying by $s(r)$ and
integrating over $r$; while the inverse map is given by varying
w.r.t. $s(r)$. However, this would be in the opposite direction from
that which led us to the states \eqref{Phi2} by imposition of the
linear operator constraints. As an example consider the action of
these operators on two particular cases of the states \eqref{Phi2},
containing the structure $s(r)$ :
\begin{eqnarray}
\hat{\mathcal{H}}_1(r) \int
s(\tilde{r})\,\rho(\tilde{r})\,\sigma(\tilde{r})\,d\tilde{r}=-s'(r)\,\rho(r)\,\sigma(r)
\neq 0 \quad \text{for arbitrary}\, s(r) \\
\hat{\mathcal{H}}_2(r) \int
s(\tilde{r})\,\chi(\tilde{r})\,d\tilde{r}=s'(r) \neq 0 \quad
\text{for arbitrary}\, s(r)\end{eqnarray}
 Thus, every foreign to the geometry structure $s(r)$ is not allowed to enter the
physical states.

Now, after the correspondence has been established, we can come to
the basic property the induced metric must have. In the case of
finite degrees of freedom the induced metric depends, up to a
conformal scaling, on the physical coordinates $q^\alpha$ by virtue
of (\ref{algekuch}). In our case, due to the dependence of the
configuration variables on the radial coordinate $r$, the above
property is not automatically satisfied; e.g. the functional
derivative $\frac{\delta}{\delta\sigma(r)}$ acting on $\Sigma^{(n)}$
will produce, upon partial integration of the $n^{th}$ derivative of
the Dirac delta function, a term proportional to $\Sigma^{(2n)}$.
Therefore, since $L$ in \eqref{Phi2} contains derivatives of
$\sigma(r)$ up to $\Sigma^{(n)}$, the above mentioned property must
be \emph{enforced}. The need for this can also be traced to the
substantially different first Poisson bracket in (\ref{alge}), which
signals a non trivial mixing between the dynamical evolution
generator $\mathcal{H}_o$ and the linear generators $\mathcal{H}_i$.\\
Thus, according to the above reasoning, in order to proceed with the
generalization of Kucha\v{r}'s method, we have to demand that:\\
\\
\textbf{Requirement:}
\emph{$L\left(\Sigma^{(0)},\ldots,\Sigma^{(n)}\right)$ must be such
that $g^{\Phi\Phi}$ becomes a general function, say
$F\left(k\,\chi(r)+\rho(r)\,L(\Sigma^{(0)},\ldots,\Sigma^{(n)})\right)$
of the integrand of $\Phi$, so that it can be considered a function
of this state:
$g^{\Phi\Phi}\overset{Assumption}{\Longleftrightarrow}F\left(k\int\,\chi(\tilde{r})d\tilde{r}+
\int\rho(\tilde{r})\,L(\Sigma^{(0)},\ldots,\Sigma^{(n)})d\tilde{r}\right)=F(\Phi)$}.\\
\\
At this point, we must emphasize that the application of the
\textbf{Requirement} in the subsequent development of our quantum
theory will result in very severe restrictions on the form of
\eqref{Phi2}. Essentially, $\chi(r)$ as well as all higher
derivatives of $\sigma(r)$ (i.e $\Sigma^{(2)}\ldots\Sigma^{(n)})$)
are eliminated from $\Phi$ (see \eqref{Phi4}, \eqref{q1,q2,q3}).
This might, at first sight, strike as odd; indeed, the common belief
is that all the derivatives of the configuration variables should
enter the physical states. However, before the imposition of both
the linear \emph{and} the quadratic constrains there are no truly
physical states. Thus, no physical states are lost by the imposition
of the \textbf{Requirement}; ultimately the only true physical
states are the solutions to \eqref{wdw}.

Having clarified the way in which we view the \textbf{Assumption}
and \textbf{Requirement} above, we now proceed to the restrictions
implied by their use.

A first consequence of the requirement that $g^{\Phi\Phi}=F\left
(k\,\chi(r)+\rho(r)\,L(\Sigma^{(0)},\ldots,\Sigma^{(n)})\right)$ is
the vanishing of $k$. This follows from (a) the property that
$g^{\Phi\Phi}$ is homogenous in the functional derivative
$\frac{\delta} {\delta \chi(r)}$, (b) that $G^{\alpha\beta}$ in
(\ref{supermetric}) does not contain any $\chi(r)$; namely
\begin{eqnarray*}
g^{\Phi\Phi}=\ldots+G^{33}\frac{\delta \Phi}{\delta
\chi(r)}\frac{\delta \Phi} {\delta \chi(r)}\leftrightarrow
g^{\Phi\Phi}=\underset{no\,\,\chi}{\underbrace{\ldots}}
+\frac{\rho}{\sigma^2}\,k^2\equiv F(k\,\chi+\rho\,L).
\end{eqnarray*}
Since $\ldots$ are terms not involving $\chi(r)$, the final
identification is possible iff $k=0$. Thus, $\Phi$ is reduced to:
\begin{equation}\label{Phi3}
\Phi=\int
d\tilde{r}\,\rho(\tilde{r})\,L\left(\Sigma^{(0)},\ldots,\Sigma^{(n)}\right).
\end{equation}

We now turn to the degree of derivatives $(n)$ of $\sigma(r)$. The
situation is similar to the corresponding case with $X^{(n)}$
considered before; again the functional derivative
$\frac{\delta}{\delta \sigma(r)}$ acting on $\Phi$ will bring a
maximum term $\Sigma^{(2n)}$ while $\frac{\delta}{\delta \rho(r)}$ a
corresponding term $\Sigma^{(2n-1)}$. More precisely
\begin{eqnarray*}
g^{\Phi\Phi}=\ldots+2\,G^{12}\frac{\delta \Phi}{\delta
\rho(r)}\frac{\delta \Phi} {\delta \sigma(r)}\,.
\end{eqnarray*}
Where the functional derivatives are:
\begin{eqnarray*}
\frac{\delta\Phi}{\delta\sigma}&=&\ldots+\int \rho\,\frac{\partial
L}{\partial \Sigma^{(n)}} \frac{\delta \Sigma^{(n)}}{\delta
\sigma}\,d\tilde{r}=\ldots+\int \rho\,\frac{\partial L} {\partial
\Sigma^{(n)}}\,\frac{1}{\rho}\,\frac{d}{d\tilde{r}}\left(\underset{n-1}
{\underbrace{\ldots}}\,\,\delta(r,\tilde{r})\right)d\tilde{r}=\\
&=&\ldots-\int \frac{d}{d\tilde{r}}\left(\frac{\partial L}{\partial
\Sigma^{(n)}}\right)
\frac{1}{\rho}\,\frac{d}{d\tilde{r}}\left(\underset{n-2}{\underbrace{\ldots}}
\,\,\delta(r,\tilde{r})\right)d\tilde{r}=\\
&=&\ldots-\int \rho\,\frac{\partial^2 L}{\partial
\left(\Sigma^{(n)}\right)^2}\,\Sigma^{(n+1)}
\,\frac{1}{\rho}\,\frac{d}{d\tilde{r}}\left(\underset{n-2}{\underbrace{\ldots}}\,\,
\delta(r,\tilde{r})\right)d\tilde{r}=\\
&=&\ldots+(-1)^n\int \rho(\tilde{r})\,\frac{\partial^2 L}{\partial
\left(\Sigma^{(n)}\right)^2}
\,\,\Sigma^{(2n)}\,\delta(r,\tilde{r})\,d\tilde{r}=\\
&=&\ldots+(-1)^n\rho\,\frac{\partial^2 L}{\partial
\left(\Sigma^{(n)}\right)^2}\,\,\Sigma^{(2n)}
\end{eqnarray*}
and
\begin{eqnarray*}
\frac{\delta\Phi}{\delta\rho}&=&\ldots+\int \rho\,\frac{\partial
L}{\partial \Sigma^{(n)}} \frac{\delta \Sigma^{(n)}}{\delta
\rho}\,d\tilde{r}=\ldots+\int \rho\,\frac{\partial L} {\partial
\Sigma^{(n)}}\,\frac{1}{\rho}\,\frac{d}{d\tilde{r}}\left(\underset{n-2}
{\underbrace{\ldots}}\,-\frac{\delta(r,\tilde{r})}{\rho(\tilde{r})^2}\,\sigma'(\tilde{r})\right)d\tilde{r}=\\
&=&\ldots+\int \rho\,\frac{\partial L}{\partial
\Sigma^{(n)}}\,\frac{1}{\rho}\,\frac{d}
{d\tilde{r}}\left(\underset{n-2}{\underbrace{\ldots}}\,-\frac{\delta(r,\tilde{r})}
{\rho(\tilde{r})}\,\Sigma^{(1)}\right)d\tilde{r}\\
&=&\ldots-\int \frac{d}{d\tilde{r}}\left(\frac{\partial L}{\partial
\Sigma^{(n)}}\right)
\frac{1}{\rho}\,\frac{d}{d\tilde{r}}\left(\underset{n-3}{\underbrace{\ldots}}\,
-\frac{\delta(r,\tilde{r})}{\rho(\tilde{r})}\,\Sigma^{(1)}\right)d\tilde{r}=\\
&=&\ldots+(-1)^{n-1}\int \frac{\partial^2 L}{\partial
\left(\Sigma^{(n)}\right)^2}
\,\,\Sigma^{(2n-1)}\,\Sigma^{(1)}\,\delta(r,\tilde{r})\,d\tilde{r}=\\
&=&\ldots+(-1)^{n-1}\frac{\partial^2 L}{\partial
\left(\Sigma^{(n)}\right)^2} \,\,\Sigma^{(2n-1)}\,\Sigma^{(1)}\,.
\end{eqnarray*}
Therefore
\begin{eqnarray*}
g^{\Phi\Phi}=\ldots-2\,\rho\,\sigma\,(-1)^{2n-1}\left(\frac{\partial^2
L}{\partial \left(\Sigma^{(n)}\right)^2}\right)^2
\Sigma^{(1)}\,\Sigma^{(2n-1)}\,\Sigma^{(2n)},
\end{eqnarray*}
where the $\ldots$ stand for all other terms, not involving
$\Sigma^{(2n)}$. Now, according to the aforementioned
\textbf{Requirement} we need this to be a general function, say
$F(\rho L)$, and for this to happen the coefficient of
$\Sigma^{(2n)}$ must vanish, i.e.
\begin{eqnarray*}
\frac{\partial^2 L}{\partial
\left(\Sigma^{(n)}\right)^2}=0\Leftrightarrow
L=L_1\left(\Sigma^{(0)},\ldots,\Sigma^{(n-1)}\right)\Sigma^{(n)}+
L_2\left(\Sigma^{(0)},\ldots,\Sigma^{(n-1)}\right).
\end{eqnarray*}
Again the term of $\Phi$ corresponding to $L_1$ is, up to a total
derivative, equivalent to a local smooth functional containing
$\Sigma^{(0)},\ldots,\Sigma^{(n-1)}$. The argument can be repeated
for $(n-1),(n-2),\ldots,2$. The case $n=1$ needs separate
consideration since, upon elimination of the linear in
$\Sigma^{(2)}$ term we are left with a local function of
$\Sigma^{(1)}$, and thus the possibility arises to meet the
\textbf{Requirement} by solving a differential equation for $L$. In
more detail, if
\begin{equation}\label{Phi4}
\Phi\equiv\int
\rho(\tilde{r})L\left(\sigma,\Sigma^{(1)}\right)d\tilde{r},
\end{equation}
$g^{\Phi\Phi}$ reads
\begin{eqnarray}
\nonumber g^{\Phi\Phi}&=&-\rho\left(L-\Sigma^{(1)}\,\frac{\partial
L}{\partial \Sigma^{(1)}}\right)
\left[L-\Sigma^{(1)}\,\frac{\partial L}{\partial
\Sigma^{(1)}}+2\,\sigma\left(\frac{\partial L}{\partial \sigma}-
\Sigma^{(1)}\,\frac{\partial^2 L}{\partial \sigma\,\partial \Sigma^{(1)}}\right)\right]+\\
\label{gphiphi1}&
&+2\,\rho\,\sigma\left(L-\Sigma^{(1)}\,\frac{\partial L}{\partial
\Sigma^{(1)}}\right)\frac{\partial^2 L}{\partial
(\Sigma^{(1)})^2}\,\Sigma^{(2)}.
\end{eqnarray}
Through the definition
\begin{equation}\label{defH}
H\equiv L-\Sigma^{(1)}\,\frac{\partial L}{\partial\Sigma^{(1)}}
\end{equation}
we obtain
\begin{eqnarray*}
\frac{\partial H}{\partial \sigma}&=&\frac{\partial L}{\partial \sigma}-\Sigma^{(1)}\,\frac{\partial^2 L}{\partial \sigma\,\partial \Sigma^{(1)}},\\
\frac{\partial H}{\partial
\Sigma^{(1)}}&=&-\Sigma^{(1)}\,\frac{\partial^2 L}{\partial
(\Sigma^{(1)})^2}\,.
\end{eqnarray*}
Thus (\ref{gphiphi1}) assumes the form
\begin{eqnarray*}
g^{\Phi\Phi}=-\rho\left(H^2+2\,\sigma\,H\,\frac{\partial
H}{\partial\sigma}+
\frac{2\,\sigma}{\Sigma^{(1)}}\,H\,\frac{\partial
H}{\partial\Sigma^{(1)}}\,\Sigma^{(2)}\right),
\end{eqnarray*}
which upon addition, by virtue of the \textbf{Assumption}, of the
surface term
\begin{eqnarray*}
A=\frac{d}{dr}\left(\int\frac{2\,\sigma}{\Sigma^{(1)}}\,H\,\frac{\partial
H}{\partial\Sigma^{(1)}}\,d\Sigma^{(1)}\right)
\end{eqnarray*}
gives
\begin{equation}\label{gphiphi2}
g^{\Phi\Phi}=-\rho\left(H^2+2\,\sigma\,H\,\frac{\partial
H}{\partial\sigma}-
\Sigma^{(1)}\frac{\partial}{\partial\sigma}\int\frac{2\,\sigma}{\Sigma^{(1)}}\,H\,\frac{\partial
H} {\partial\Sigma^{(1)}}\,d\Sigma^{(1)}\right).
\end{equation}
Since in the last expression we have only a multiplicative
$\rho(r)$, it is obvious that the \textbf{Requirement}
\begin{eqnarray*}
g^{\Phi\Phi}=F(\rho\,L)
\end{eqnarray*}
can be satisfied only by
\begin{equation}\label{gphiphi3}
g^{\Phi\Phi}=-\kappa\,\rho\,L,
\end{equation}
with $g^{\Phi\Phi}$ given by (\ref{gphiphi2}). Upon differentiation
of this equation with respect to $\Sigma^{(1)}$ we get
\begin{eqnarray*}
-\frac{\partial}{\partial\sigma}\int\frac{2\,\sigma}{\Sigma^{(1)}}\,H\,\frac{\partial
H} {\partial\Sigma^{(1)}}=\kappa\,\frac{\partial
L}{\partial\Sigma^{(1)}}\,.
\end{eqnarray*}
Multiplying the last expression by $\Sigma^{(1)}$ and subtracting it
from (\ref{gphiphi3}) we end up with the autonomous necessary
condition for $H(\sigma,\,\Sigma^{(1)})$:
\begin{eqnarray*}
H\left(H+2\,\sigma\,\frac{\partial
H}{\partial\sigma}-\kappa\right)=0,
\end{eqnarray*}
where (\ref{defH}) was also used. The above equation can be readily
integrated giving
\begin{eqnarray*}
H&=&0, \\
H&=&\kappa+\frac{a(\Sigma^{(1)})}{\sqrt{\sigma}},
\end{eqnarray*}
where $a(\Sigma^{(1)})$ is an arbitrary function of its argument.
The first possibility gives according to (\ref{defH})
$L=\lambda\,\Sigma^{(1)}$ which, however, contributes to $\Phi$ a
surface term, and can thus be ignored. Inserting the second solution
into (\ref{defH}) we construct a partial differential equation for
$L$, namely
\begin{eqnarray*}
L-\Sigma^{(1)}\,\frac{\partial
L}{\partial\Sigma^{(1)}}=\kappa+\frac{a(\Sigma^{(1)})}{\sqrt{\sigma}},
\end{eqnarray*}
which upon integration gives
\begin{eqnarray*}
L=\kappa-\frac{\Sigma^{(1)}}{\sqrt{\sigma}}\int\frac{a(\Sigma^{(1)})}{{\Sigma^{(1)}}^2}\,\,d\Sigma^{(1)}+
c_1(\sigma)\,\Sigma^{(1)}\,.
\end{eqnarray*}
Since this form of $L$ emerged as a necessary condition, it must be
inserted (along with $H$) in (\ref{gphiphi3}). The result is that
$c_1(\sigma)=0$. Thus $L$ reads
\begin{equation}\label{L}
L=\kappa-\frac{\Sigma^{(1)}}{\sqrt{\sigma}}\int\frac{a(\Sigma^{(1)})}{{\Sigma^{(1)}}^2}\,\,d\Sigma^{(1)}\,.
\end{equation}
By assuming that the $\Sigma^{(1)}$--dependent part of $L$ equals
$b(\Sigma^{(1)})$, i.e.
\begin{eqnarray*}
-\Sigma^{(1)}\int\frac{a(\Sigma^{(1)})}{{\Sigma^{(1)}}^2}\,\,d\Sigma^{(1)}=b(\Sigma^{(1)}),
\end{eqnarray*}
we get, upon a double differentiation with respect to
$\Sigma^{(1)}$, the ordinary differential equation
\begin{eqnarray*}
-\frac{a\,'(\Sigma^{(1)})}{\Sigma^{(1)}}=b\,''(\Sigma^{(1)})
\end{eqnarray*}
with solution
\begin{eqnarray*}
a(\Sigma^{(1)})=b(\Sigma^{(1)})+\kappa_1-\Sigma^{(1)}\,b\,'(\Sigma^{(1)}),
\end{eqnarray*}
where $\kappa_1$ is a constant. Substituting this equation into
(\ref{L}) and performing a partial integration we end up with
\begin{equation}\label{L1}
L=\kappa+\frac{\kappa_1}{\sqrt{\sigma}}+\frac{b(\Sigma^{(1)})}{\sqrt{\sigma}}\,.
\end{equation}
$\kappa$, $\kappa_1$ and $b(\Sigma^{(1)})$ being completely
arbitrary and to our disposal; the two simpler choices
$\kappa_1=0,\,b(\Sigma^{(1)})=0$ and $\kappa=0,\,b(\Sigma^{(1)})=0$
lead respectively to the following two basic local smooth
functionals:
\begin{equation}\label{q1,q2}
q^1=\int d\tilde{r}\rho(\tilde{r}), \qquad q^2=\int
d\tilde{r}\frac{\rho(\tilde{r})} {\sqrt{\sigma(\tilde{r})}}\,.
\end{equation}
The next simplest choice $\kappa=0,\,\kappa_1=0$ and
$b(\Sigma^{(1)})$ arbitrary leads to a generic $q^3=\int
d\tilde{r}\rho(\tilde{r})\,
\frac{b(\Sigma^{(1)})}{\sqrt{\sigma(\tilde{r})}}$. However, it can
be proven that, for any choice of $b(\Sigma^{(1)})$, the
corresponding renormalized induced metric
\begin{eqnarray*}
g^{AB}=G^{\alpha\beta}\frac{\delta q^A}{\delta x^\alpha}\frac{\delta
q^B}{\delta x^\beta} \qquad where \quad A,B=1,2,3
\end{eqnarray*}
is singular. The calculation of $g^{AB}$ gives:
\begin{eqnarray*}
g^{11}&=&G^{\alpha\beta}\frac{\delta q^1}{\delta
x^\alpha}\frac{\delta q^1}{\delta x^\beta}=
-\rho\overset{Assumption}{\Longleftrightarrow}g_{ren}^{11}=-q^1,\\
g^{12}&=&G^{\alpha\beta}\frac{\delta q^1}{\delta
x^\alpha}\frac{\delta q^2}{\delta x^\beta}=
-\frac{\rho}{2\sqrt{\sigma}}\overset{Assumption}{\Longleftrightarrow}g_{ren}^{12}=-\frac{q^2}{2}\,,\\
g^{22}&=&G^{\alpha\beta}\frac{\delta q^2}{\delta
x^\alpha}\frac{\delta q^2}{\delta x^\beta}=
0\overset{Assumption}{\Longleftrightarrow}g_{ren}^{22}=0,\\
g^{13}&=&G^{\alpha\beta}\frac{\delta q^1}{\delta
x^\alpha}\frac{\delta q^3}{\delta x^\beta}=
\rho\left(-\frac{b}{2\,\sqrt{\sigma}}+\frac{\Sigma^{(1)}}{2\,\sqrt{\sigma}}\,b\,'+
\sqrt{\sigma}\,\Sigma^{(2)}\,b\,''\right)\overset{Assumption}{\Longleftrightarrow}\\
g_{ren}^{13}&=&\int
dr\rho\left(-\frac{b}{2\,\sqrt{\sigma}}+\frac{\Sigma^{(1)}}{2\,\sqrt{\sigma}}\,b\,'+
\sqrt{\sigma}\,\Sigma^{(2)}\,b\,''\right)-\int dr\frac{d}{dr}\left(\int d\Sigma^{(1)}\,\sqrt{\sigma}\,b\,''\right)=\\
&=&-\int dr\rho\,\frac{b}{2\,\sqrt{\sigma}}=-\frac{q^3}{2}\,,
\end{eqnarray*}
\begin{eqnarray*}
g^{23}&=&G^{\alpha\beta}\frac{\delta q^2}{\delta
x^\alpha}\frac{\delta q^3}{\delta x^\beta}=
\rho\,\Sigma^{(2)}\,b\,''=\frac{d}{dr}\,b\,'\overset{Assumption}{\Longleftrightarrow}g_{ren}^{23}=0,\\
g^{33}&=&G^{\alpha\beta}\frac{\delta q^3}{\delta
x^\alpha}\frac{\delta q^3}{\delta x^\beta}=
2\,\rho\left(b-\Sigma^{(1)}\,b'\right)\Sigma^{(2)}\,b\,''\overset{Assumption}{\Longleftrightarrow}\\
g_{ren}^{33}&=&2\int
dr\rho\left(b-\Sigma^{(1)}\,b'\right)\Sigma^{(2)}\,b\,''- 2\int
dr\,\frac{d}{dr}\left[\int
d\Sigma^{(1)}\left(b-\Sigma^{(1)}\,b'\right)\,b\,''\right]=0,
\end{eqnarray*}
where by $'$ we denote differentiation with respect to
$\Sigma^{(1)}$. Thus the renormalized induced metric reads
\begin{eqnarray*}
g^{AB}_{ren}=-\frac{1}{2}
\begin{pmatrix}
\displaystyle{2\,q^1} & \displaystyle{q^2} & \displaystyle{q^3} \cr
& \cr \displaystyle{q^2} & 0  & 0 \cr & \cr \displaystyle{q^3} & 0 &
0 \cr
\end{pmatrix}.
\end{eqnarray*}
Effecting the transformation
$(\tilde{q}^1,\,\tilde{q}^2,\,\tilde{q}^3)=\left(q^1,\,q^2,\,\ln
\frac{q^3}{q^2}\right)$ we bring $g^{AB}_{ren}$ into a manifestly
degenerate form:
\begin{eqnarray*}
g^{AB}_{ren}=-\frac{1}{2}
\begin{pmatrix}
\displaystyle{2\,q^1} & \displaystyle{q^2} & \displaystyle{0} \cr &
\cr \displaystyle{q^2} & 0  & 0 \cr & \cr \displaystyle{0} & 0 & 0
\cr
\end{pmatrix}.
\end{eqnarray*}
So, it seems as though the relevant part of the renormalized metric
is described by the upper $2\times2$ block. This fact is consistent
with the form of the renormalized potential $V=2\,\Lambda\,q^1$
which indeed does not contain any $\Sigma^{(1)}$ term.

However, this is not the end of our investigation for a suitable
space of state vectors: the argument leading to $q^1,\,q^2$ depends
upon the original choice of \underline{one} initial candidate smooth
scalar functional (\ref{Phi4}); to complete the search we must close
the circle by starting with the two already secured smooth
functionals $(q^1,\,q^2)$, and a \underline{third} of the general
form
\begin{eqnarray*}
q^3=\int dr\,\rho\,L(\Sigma^{(1)}),
\end{eqnarray*}
since the $\sigma$ dependence has already been fixed to either $1$
or $\frac{1}{\sqrt{\sigma}}$\,. The calculation of the, related to
$q^3$, components of the induced metric $g^{AB}$ gives:
\begin{eqnarray*}
g^{13}&=&\rho\left(-L+\Sigma^{(1)}\,L'+\sigma\,\Sigma^{(2)}\,L''\right)\overset{Assumption}{\Longleftrightarrow}\\
g_{ren}^{13}&=&\int
dr\rho\left(-L+\Sigma^{(1)}\,L'+\sigma\,\Sigma^{(2)}\,L''\right)-
\int dr\frac{d}{dr}\left(\int d\Sigma^{(1)}\,\sigma\,L''\right)=-\int dr\rho\,L=\\&=&-q^3,\\
g^{23}&=&\rho\left(-\frac{L}{2\,\sqrt{\sigma}}+\frac{\Sigma^{(1)}}{2\,\sqrt{\sigma}}\,L'+\sqrt{\sigma}\,\Sigma^{(2)}L''\right)
\overset{Assumption}{\Longleftrightarrow}\\
g_{ren}^{23}&=&\int
dr\rho\left(-\frac{L}{2\,\sqrt{\sigma}}+\frac{\Sigma^{(1)}}{2\,\sqrt{\sigma}}\,L'+
\sqrt{\sigma}\,\Sigma^{(2)}\,L''\right)-\int dr\frac{d}{dr}\left(\int d\Sigma^{(1)}\,\sqrt{\sigma}\,L''\right)=\\
&=&-\int dr\rho\,\frac{L}{2\,\sqrt{\sigma}}\overset{Assumption}{=}-\frac{q^2\,q^3}{2\,q^1}\,,\\
g^{33}&=&-\rho\left(L-\Sigma^{(1)}L'\right)^2+2\,\rho\,\sigma\left(L-\Sigma^{(1)}L'\right)\Sigma^{(2)}L''.
\end{eqnarray*}
By following the procedure presented between (\ref{gphiphi1}) and
(\ref{gphiphi2}) we end up with the expression
\begin{eqnarray*}
g^{33}=-\rho\left[\left(L-\Sigma^{(1)}L'\right)^2-
\Sigma^{(1)}\int\frac{d\Sigma^{(1)}}{\Sigma^{(1)}}\,\frac{\partial}{\partial\Sigma^{(1)}}\left(L-
\Sigma^{(1)}L'\right)^2\right],
\end{eqnarray*}
the expression inside the square brackets being a generic function
of $\Sigma^{(1)}$ and therefore, also of $L$: let this function be
parameterized as
$L\left(\Sigma^{(1)}\right)^2-\frac{4\,F[L\left(\Sigma^{(1)}\right)]^2}{3\,F'[F[L\left(\Sigma^{(1)}\right)]]^2}$\,;
this ``peculiar'' parametrization of the arbitrariness in
$L\left(\Sigma^{(1)}\right)$ has been chosen in order to facilitate
the subsequent proof that this freedom is a pure general coordinate
transformation (gct) of the induced re-normalized metric. Indeed,
let us first take the simplest non trivial choice
$L\left(\Sigma^{(1)}\right)\equiv{\Sigma^{(1)}}^2$ which results in
the re-normalized metric
\begin{equation}\label{renmet}
g^{AB}_{ren}=-\frac{1}{2}
\begin{pmatrix}
\displaystyle{2\,q^1} & \displaystyle{q^2} & \displaystyle{2\,q^3}
\cr & \cr \displaystyle{q^2} & 0  &
\displaystyle{\frac{q^2q^3}{q^1}} \cr & \cr \displaystyle{2\,q^3} &
\displaystyle{\frac{q^2q^3}{q^1}} &
-\displaystyle{\frac{2\,(q^3)^2}{3\,q^1}} \cr
\end{pmatrix}, \qquad
{g_{AB}}_{ren}=\frac{1}{2}
\begin{pmatrix}
\displaystyle{\frac{3}{2\,q^1}} & -\displaystyle{\frac{4}{q^2}} &
-\displaystyle{\frac{3}{2\,q^3}} \cr & \cr
-\displaystyle{\frac{4}{q^2}} &
\displaystyle{\frac{8\,q^1}{(q^2)^2}} & 0 \cr & \cr
-\displaystyle{\frac{3}{2\,q^3}} & 0 &
\displaystyle{\frac{3\,q^1}{2(q^3)^2}} \cr
\end{pmatrix}.
\end{equation}
Considering a generic $L\left(\Sigma^{(1)}\right)$, i.e. $x^3=\int
dr\,\rho\,L(\Sigma^{(1)})$ (along with (\ref{q1,q2})) we are led to
\begin{eqnarray*}
g^{13}=-x^3, \qquad g^{23}=-\frac{q^2\,x^3}{2\,q^1}
\end{eqnarray*}
and
\begin{eqnarray*}
g^{33}=-\rho\,\left[L^2-\frac{4\,F[L]^2}{3\,F'[F[L]]^2}\right]&=&
-\frac{(\rho\,L)^2}{\rho}+\frac{4\,\rho\,F[\frac{\rho\,L}{\rho}]^2}{3\,F'[F[\frac{\rho\,L}{\rho}]]^2}\overset{Assumption}{\Longleftrightarrow}
\\ \nonumber g_{ren}^{33}=-\frac{(x^3)^2}{q^1}+\frac{4\,q^1\,F[\frac{x^3}{q^1}]^2}{3\,F'[F[\frac{x^3}{q^1}]]^2}\,
\end{eqnarray*}
Remarkably enough, the new induced re-normalized metric can be put
in gct equivalence with the metric (\ref{renmet}) through the
transformation
\begin{eqnarray*}
(q^1,q^2,x^3)=(q^1,q^2,q^1\,F^{-1}(\frac{q^3}{q^1})),
\end{eqnarray*}
with $F^{-1}$ denoting the function inverse to $F,$ i.e.
$F^{-1}(F(x))\equiv\,x$ .

We can therefore consider, without loss of generality, the reduced
re-normalized manifold to be parameterized by the following three
smooth scalar functionals:
\begin{equation}\label{q1,q2,q3}
q^1=\int d\tilde{r}\rho(\tilde{r}), \qquad q^2=\int
d\tilde{r}\frac{\rho(\tilde{r})} {\sqrt{\sigma(\tilde{r})}}, \qquad
q^3=\int d\tilde{r}\,\frac{\sigma'(\tilde{r})^2}{\rho(\tilde{r})}\,.
\end{equation}

Any other functional, say $q^4=\int
d\tilde{r}\,\rho(\tilde{r})\,L(\sigma(\tilde{r}),\Sigma^{(1)}(\tilde{r}))$,
can be considered as a function of $q^1,q^2,q^3$; indeed, since the
scalar functions appearing in the integrands of $q^2, q^3$ form a
base in the space of $\sigma, \Sigma^{(1)}$, we can express the
generic $L$ in $q^4$ as
$F(\frac{\rho}{\rho\,\sqrt{\sigma}},\frac{\rho\,
\Sigma^{(1)2}}{\rho}))$, which (through the \textbf{Assumption})
gives $q^4=q^1 F(\frac{q^2}{q^1},\frac{q^3}{q^1})$.

 The geometry of this space is described by the induced
re-normalized metric (\ref{renmet}). Any function
$\Psi(q^1,q^2,q^3)$ on this manifold is of course annihilated by the
quantum linear constraints, i.e.
\begin{eqnarray*}
\hat{\mathcal{H}}_1\Psi(q^1,q^2,q^3)=\frac{\partial\Psi(q^1,q^2,q^3)}{\partial
q^1}\,\hat{\mathcal{H}}_1\,q^1+
\frac{\partial\Psi(q^1,q^2,q^3)}{\partial
q^2}\,\hat{\mathcal{H}}_1\,q^2+
\frac{\partial\Psi(q^1,q^2,q^3)}{\partial
q^3}\,\hat{\mathcal{H}}_1\,q^3=0
\end{eqnarray*}
\begin{eqnarray*}
\hat{\mathcal{H}}_2\Psi(q^1,q^2,q^3)=\frac{\partial\Psi(q^1,q^2,q^3)}{\partial
q^1}\,\hat{\mathcal{H}}_2\,q^1+
\frac{\partial\Psi(q^1,q^2,q^3)}{\partial
q^2}\,\hat{\mathcal{H}}_2\,q^2+
\frac{\partial\Psi(q^1,q^2,q^3)}{\partial
q^3}\,\hat{\mathcal{H}}_2\,q^3=0
\end{eqnarray*}
since the derivatives with respect to $r$ are transparent to the
partial derivatives of $\Psi$ (which are, just like the $q^A$'s,
r-numbers).

The final restriction on the form of $\Psi$ will be obtained by the
imposition of the quantum analog of the quadratic constraint
$\mathcal{H}_o$. According to the above exposition we postulate that
the quantum gravity of the geometries given by (\ref{canmetric}),
(\ref{g metric}) will be described by the following partial
differential equation (in terms of the $q^A$'s)
\begin{equation}\label{wdw}
\hat{\mathcal{H}}_o\Psi\equiv [-\frac{1}{2}\,\Box_{c}
+V_{ren}]\,\Psi(q^1,q^2,q^3)=0
\end{equation}
with
\begin{equation}\label{conf Lap}
\Box_c=\Box+\frac{d-2}{4\,(d-1)}\,R
\end{equation}
being the conformal Laplacian based on  $g_{AB\, ren}$,  $R$ the
Ricci scalar, and $d$ the dimensions of $g_{AB\, ren}$. Metric
(\ref{renmet}) is conformally flat with Ricci scalar
$R=\frac{1}{2\,q^1}$, and its dimension is $d=3$. The re-normalized
form of the potential (\ref{potential}) offers us the possibility to
introduce, in a dynamical way, topological effects into our wave
functional: Indeed, under our \textbf{Assumption}, the first term
becomes $2\,\Lambda\,q^1$ while the second, being a total
derivative, becomes $ A_T \equiv
\frac{\sigma^\prime}{\rho}\mid^\beta_\alpha \,(if\,\alpha<r<\beta)
$. In the spirit previously explained we should drop this term,
however one could keep it, thus arriving at $V_{ren}=2\,\Lambda\,
q^1+A_T$ and the Wheeler-deWitt equation is finally given as

\begin{eqnarray}\label{WdW}
&&2\,q^1\,\Lambda\,\Psi(q^1,q^2,q^3)+A_T\,\Psi(q^1,q^2,q^3)-\frac{1}{32\,q^1}\Psi(q^1,q^2,q^3)+
\frac{q^3}{12\,q^1}\frac{\partial\Psi(q^1,q^2,q^3)}{\partial q^3}+\nonumber\\
&&\frac{q^2}{8\,q^1}\frac{\partial\Psi(q^1,q^2,q^3)}{\partial q^2}+
\frac{3}{4}\frac{\partial\Psi(q^1,q^2,q^3)}{\partial q^1}+
\frac{q^2q^3}{2\,q^1}\frac{\partial^2\Psi(q^1,q^2,q^3)}{\partial
q^2\,\partial q^3}+ q^3\,\frac{\partial^2\Psi(q^1,q^2,q^3)}{\partial
q^1\,\partial q^3}+\nonumber\\
&&\frac{q^2}{2}\frac{\partial^2\Psi(q^1,q^2,q^3)}{\partial
q^1\,\partial q^2}+
\frac{q^1}{2}\frac{\partial^2\Psi(q^1,q^2,q^3)}{\partial\left(q^1\right)^2}-
\frac{\left(q^3\right)^2}{6\,q^1}\frac{\partial^2\Psi(q^1,q^2,q^3)}{\partial\left(q^3\right)^2}=0.
\end{eqnarray}
The change to new coordinates $(x^1,\,x^2,\,x^3)$ described by
\begin{eqnarray*}
(q^1,\,q^2,\,q^3)=\left(e^{x^2},\,e^{\frac{1}{2}(x^1+x^2)},\,e^{x^2+\frac{2}{\sqrt{3}}x^3}\right)
\end{eqnarray*}
transforms the metric into the manifestly conformally flat form
$diag\{e^{x^2},-e^{x^2},\,-e^{x^2}\}$ and brings (\ref{WdW}) into
the form
\begin{eqnarray}\label{WdWnew}
&&2\,e^{2\,x^2}\,\Lambda\,\Psi(x^1,x^2,x^3)+A_T\,e^{x^2}\,\Psi(x^1,x^2,x^3)-\frac{1}{32}\,\Psi(x^1,x^2,x^3)+
\frac{1}{4}\frac{\partial\Psi(x^1,x^2,x^3)}{\partial x^2}-\nonumber\\
&&\frac{1}{2}\frac{\partial^2\Psi(x^1,x^2,x^3)}{\partial\left(x^1\right)^2}+
\frac{1}{2}\frac{\partial^2\Psi(x^1,x^2,x^3)}{\partial\left(x^2\right)^2}+
\frac{1}{2}\frac{\partial^2\Psi(x^1,x^2,x^3)}{\partial\left(x^3\right)^2}=0.
\end{eqnarray}
This equation is readily solved by the method of separation of
variables: assuming $\Psi(x^1,x^2,x^3)=X^1(x^1)\,X^2(x^2)\,X^3(x^3)$
and dividing (\ref{WdWnew}) by $\Psi$ we get the three ordinary
differential equations: \bsub
\begin{eqnarray}
&&\frac{1}{2\,X^1(x^1)}\frac{d\,^2 X^1(x^1)}{d\left(x^1\right)^2}+\frac{1}{32}=m+n, \\
&&\frac{1}{2\,X^2(x^2)}\frac{d\,^2 X^2(x^2)}{d\left(x^2\right)^2}+
\frac{1}{4\,X^2(x^2)}\frac{d X^2(x^2)}{d x^2}+2\,e^{2\,x^2}\,\Lambda+A_T\,e^{x^2}=n,\\
&&\frac{1}{2\,X^3(x^3)}\frac{d\,^2 X^3(x^3)}{d\left(x^3\right)^2}=m,
\end{eqnarray}
\esub where $m$ and $n$ are separation constants. Their solutions
for $A_T=0$ are: \bsub
\begin{eqnarray}\label{solwdw}
X^1(x^1)&=&c_1\,e^{\frac{1}{4}\sqrt{32\,m+32\,n-1}\,\,x^1}+c_2\,e^{-\frac{1}{4}\sqrt{32\,m+32\,n-1}\,\,x^1}\,, \\
X^2(x^2)&=&c_3\,e^{-x^2/4}\,J_{-\frac{1}{4}\sqrt{32\,n+1}}\left(2\,e^{x^2}\sqrt{\Lambda}\right)
+c_4\,e^{-x^2/4}\,J_{\frac{1}{4}\sqrt{32\,n+1}}\left(2\,e^{x^2}\sqrt{\Lambda}\right)\\
X^3(x^3)&=&c_5\,e^{\sqrt{2\,m}\,\,x^3}+c_6\,e^{-\sqrt{2\,m}\,\,x^3}\,
,
\end{eqnarray}\esub
where $J_{\pm
\frac{1}{4}\sqrt{32\,n+1}}\left(2\,e^{x^2}\sqrt{\Lambda}\right)$
Bessel functions of the first kind and of non-integral order.

\section{Discussion}
We have considered the canonical analysis and subsequent
quantization of the (2+1)-dimensional action of pure gravity plus a
cosmological constant term, under the assumption of the existence of
one Killing vector field. The implementation of the Dirac algorithm
for this action results, at the classical level, in two linear
(momentum) and one quadratic (Hamiltonian) first class constraints.
The first linear constraint (\ref{H1}) is shown to correspond to
arbitrary changes of the radial coordinate. The second linear
constraint (\ref{H2}) owes its existence to the $G_1$ symmetry
imposed, a fact that is by itself worth mentioning. The quadratic
constraint (\ref{Ho}) is, as usual, the generator of the time
evolution (using the classical equation of motion, see pp. 21 of
\cite{Carlip1}). To avoid an ill-defined action of the quantum
analogues of the linear constraints, we adopt as our initial
collection of state vectors all smooth (integrals over the radial
coordinate $r$) functionals. The first quantum linear constraint
entails a reduction of this collection to all smooth scalar
functionals (\ref{scalfun}). The subsequent imposition of the second
quantum linear constraint further reduces these states to
(\ref{Phi2}). At this stage the need emerges to somehow obtain,
through the midi-superspace metric (\ref{supermetric}), an induced
metric (\ref{physical}) whose components are given in terms of the
same states. This leads us to firstly adopt a particular (formal)
re-normalization prescription (see \textbf{Assumption} pp. 11) and
secondly impose the \textbf{Requirement}. As a result, the final
collection of state vectors is reduced to the three unique smooth
scalar functionals (\ref{q1,q2,q3}). The quantum analogue of the
kinetic part of (\ref{Ho}) is then realized as the conformal
Laplace-Beltrami operator based on the induced re-normalized metric
(\ref{renmet}), resulting in the Wheeler-DeWitt equation
(\ref{WdW}). Effecting an appropriate change of variables the
equation is made separable and, subsequently, completely integrated.

We now come to two issues we deem worth-wile discussing:

The first has to do with the apparent absence of the quantum
analogue of the classical Poisson algebra (\ref{alge}). It seems to
us that the primary purpose of searching for a (self-adjoint)
representation of this algebra on a Hilbert space is to secure,
through Frobenius' Theorem, the consistency of the quantum theory
emanating from the chosen operator constraints ({\ref{qulincon}),
(\ref{qulincon2}) and (\ref{wdw}). But this aim is superseded by the
finding of the common kernel, i.e. the solutions  (\ref{solwdw}).
Furthermore, if, after the issue of the measure is resolved, the
Hilbert space is to be composed out of these states, the algebra of
the operator constraints will be reduced to an Abelian one.

The second concerns our choice of following Dirac's Proposal to
implement the first class constraints (\ref{H1}),(\ref{H2}) and
(\ref{Ho}) as operator conditions annihilating the wave functional,
rather than "imposing" them at the classical level, as is the case
for the vast majority of relevant works in $2+1$ gravity. Within
Dirac's Theory for constraint systems the only correct way we are
aware of to impose the first class constraints at the classical
level is to choose a ``gauge", i.e. to select a phase-space function
for each first class constraint so that constraints plus ``gauge"
fixing conditions become second class: then and only then one is
allowed to solve them all, at the very important expense of being
obliged there-afterwards to use Dirac rather than Poisson brackets.
Since the construction of these brackets makes use of the matrix
formed by the second class constraints, it is obvious that one will,
in general, be carrying to the subsequent quantization procedure
properties of the choice made. In such a situation one is never
certain of how and/or to what extent the ``gauge" fixing chosen will
infiltrate and affect the emanating quantum theory, especially if
the ``gauge" involved is so immense and complicated as the group of
space-time coordinate transformations. This constitutes our primary
motivation for following Dirac's proposal which we interpret as an
elimination of the ``gauge" freedom at the quantum level. The fact
that in 2+1 dimensions it seems more easy to classically separate
the ``gauge" from the ``true" degrees of freedom does not at all
diminish the strength of this motivation, much more in view of the
fact that our method is meant to be applicable to spherically
symmetric 3+1 geometries as well.

Generally (and somewhat loosely) speaking, the point of the exercise
as we see it is, at a first stage, to assign a unique number between
0 and 1 to each and every geometry (\ref{canmetric})-(\ref{g
metric}), in a way that is independent of the coordinate system used
to represent the metric. Of course, at the present status of things
we cannot do this, since the following two problems remain to be
solved: i) render finite the three smooth functionals
(\ref{q1,q2,q3}) and ii) select an appropriate inner product.

The first will need a final regularization of $q^1,q^2,q^3$, but
most probably, the detailed way to do this will depend upon the
particular geometry under consideration. For example, it is obvious
that for the metric (\ref{BTZ}) three segments of the range
($0,\infty$) of the radial coordinate have to be separately
considered, while for the metrics (\ref{cosmo1}), (\ref{cosmo2}) one
segment (the entire range) is enough.

For the second, a natural choice would be the determinant of the
induced re-normalized metric, although the problem with the positive
definiteness may dictate another choice.

An analogous treatment of the (3+1)-dimensional spherically
symmetric configurations can be carried through, a task that we have
already under active consideration.
\newpage
\section*{Acknowledgments}
One of the authors (T. C.) acknowledges pleasant and fruitful
discussions with Ass. Prof. C. Chiou and Prof. C. N. Ktorides.
Another author (G. O. Papadopoulos) is a Killam Postdoctoral Fellow
and acknowledges the relevant support from the Killam Foundation.

\end{document}